\documentclass[12pt]{iopart}
\usepackage{iopams}

\usepackage{bm}
\usepackage{epsfig}

\begin{document}

\title{Effect of core-valence intra-atomic quadrupolar interaction in
  resonant x-ray scattering at the Dy M$_{4,5}$ edges in DyB$_2$C$_2$}

\author{Javier Fern\'{a}ndez-Rodr\'{\i}guez}
\author{Alessandro Mirone}

\address{European Synchrotron Radiation Facility, BP 220, 38043  Grenoble Cedex, France}

\author{Urs Staub}

\address{Swiss Light Source, Paul Scherrer Institut, 5232 Villigen PSI, Switzerland}

\date{\today}

\begin{abstract}
The dependence with energy of the resonant soft x-ray Bragg diffraction
intensity in DyB$_2$C$_2$  for the $(00\frac{1}{2})$ reflection at the Dy M$_{4,5}$
edges have been calculated by using an atomic multiplet hamiltonian
including the effect of crystal field and introducing an intra-atomic
quadrupolar interaction between the 3d core and 4f valence shell. These
calculations are compared with the experimental results (Mulders et
al., J. Phys.: Condens. Matter 18 (2006) 11195) in the
antiferroquadrupolar and antiferromagnetic phases of DyB$_2$C$_2$. We
reproduce all the features appearing $(00\frac{1}{2})$ reflection energy profile
in the antiferroquadrupolar ordered phase, and we reproduce the
behaviour of the resonant x-ray scattering intensity at different
energies in the vicinity of the Dy M$_5$ edge when the temperature is
lowered within the antiferromagnetic phase.

\end{abstract}

\maketitle

\section{Introduction}

The charge, orbital, and spin degrees of freedom of the few electrons
in the valence states of a material play an important role in its
electronic properties. Resonant x-ray scattering (RXS), enhanced by the
brightness, tunability, and high degree of polarization available at
x-ray synchrotron sources, is an effective technique to measure these
microscopic variables. In RXS, by studying intensities at
space-group-forbidden reflections, dipolar and quadrupolar order
parameters can be studied
~\cite{PRBLovesey2004,Fernandez2005,McMorrow2001,Walker2006,Lovesey2005}. Only recently has
it become possible to access such ordering phenomena by means of
resonant x-ray diffraction. Higher multipolar orderings can be studied
by resonant x-ray diffraction: octupoles ~\cite{Lovesey2007a}, hexadecapoles
~\cite{Tanaka2004PRB,Fernandez2008}, anapoles ~\cite{Lovesey2007b, Fernandez2009}.
Also non-resonant x-ray diffraction can give information on higher
order multipolar orderings ~\cite{Tanaka2004EPL}.

DyB$_2$C$_2$ shows the highest antiferroquadrupole (AFQ) transition with
$T_Q = 24.7$~K ~\cite{9PRL}. The ordering of quadrupoles in DyB$_2$C$_2$ has
been studied intensively with resonant x-ray scattering
~\cite{Tanaka99,Hirota2000,Tanaka2004PRB,Matsumura2002} and neutron diffraction
in magnetic fields~\cite{Yamauchi2003,Zaharko2004}. At the AFQ ordering
temperature $T_Q$ the space group symmetry of the material is reduced
from P4/mbm to P4$_2$/mnm~\cite{Lovesey2001} with a doubling of the unit cell
along the c-axis. Below $T_N = 15.3$~K magnetic order appears, which has
been observed in neutron diffraction ~\cite{Yamauchi2003,Zaharko2004}.
Inelastic neutron scattering has been used to study the magnetic
dipolar and orbital fluctuations in this compound ~\cite{Staub2005}. Mulders
et al. ~\cite{Mulders2006} report isotropic absorption measurements and soft
x-ray resonant diffraction at the $(00\frac{1}{2})$ space group forbidden
reflection at the Dy M$_{4,5}$ edges in the antiferroquadrupolar ordered
(AFQ) and antiferromagnetic phase (AFM). A possible charge scattering
contribution is ruled out, as then s and p incident spectra would
differ by orders of magnitude and would have a completely different
shape in energy as the Bragg angle is close to 45 degrees. In order to
justify the shape of the resonant diffraction energy profile, Mulders
et al. ~\cite{Mulders2006} use a simple model for describing resonant x-ray
scattering in terms of single oscillators for each of the $\bar{M}$ quantum
numbers for the $\bar{J} = 5/2, 3/2$ core holes at the $M_{5,4}$ edges. The
degeneration of the resonant oscillators would be splited due to a
Coulomb intra-atomic quadrupolar interaction, given by 
$[3 \bar{M}^2 - \bar{J} (\bar{J}+1)]
Q_{\bar{J}}$, similarly to Mossbauer spectroscopy, and which would be
produced by the ordered quadrupolar moment in the 4f shell. A similar
core-valence interaction between the valence octupole moment and the
core hole was used in NpO$_2$ ~\cite{Lovesey2003}. In this paper we explore
the effect of such dependence of the core-hole interaction, going
beyond the analysis done in ~\cite{Mulders2006} by using a full atomic
multiplet Hamiltonian including the effect of crystal field.

\section{Calculation of the absorption and RXS spectra}

We calculate the resonant x-ray scattering factor for the dipolar
transition 4f$^9$ $\rightarrow$ 3$d^9$4f$^10$ by making use of the
program Hilbert++ ~\cite{Mirone2006,Mirone2007}. The program starts from a
model Hamiltonian accounting for multiplets and hybridization, written
in terms of creation and destruction operators, and applies Lanczos
tridiagonalization to the Hilbert space spanned by the electronic
degrees of freedom of the absorber ion and its nearest neighbours to
find the ground state $| g \rangle$. The resonant x-ray scattering tensor is
calculated as
\begin{equation}
\langle  g |\varepsilon' \mathbf{D} | \frac{  {1} } {  {\omega - H_e +
    i \gamma} }  | \varepsilon \mathbf{D} |  g \rangle 
\end{equation}
where $H_e$ is the hamiltonian of the excited state, $D$ is the dipole
operator and $\Gamma$ is the lifetime broadening. We use a model
Hamiltonian that includes the atomic multiplets and the crystal field
interaction with the neighbouring ions,
\begin{equation}
H = H_{\mathrm{atomic}} + H_{\mathrm{CF}} .
\end{equation}
Cowan's atomic multiplet program provides ab initio Hartree-Fock values
of the radial Coulomb Slater integrals and the spin-orbit interactions
for an isolated ion. In order to take into account the screening
effects present in the real system with respect to the atomic picture
we scale down all the Slater integrals to 75\% of their atomic values.
To model the interaction with the neighbouring ions, we construct a
crystal field term simulating the effect of hybridization with the
carbon first neighbour ions,
\begin{equation}
H_{\mathrm{CF}} =\sum_{b} V_{\sigma , b}  f^{+}_{{\tilde  z} ^{3}} f_{{\tilde  z} ^{3}}
\end{equation}
where $V_{\sigma}$ is the energy displacement in the Dy $f_{z^3}$
orbital produced by a ligand ion along the z axis and
$f_{z^3}^\dagger/f_{z^3}$ denote creation/destruction operators in the
f shell of the Dy ion being the local z axis orientated along the bond
direction of each of the neighbouring atoms. This term is summed over
the bonding atoms (we consider 8 nearest neighbours). The parameter
$V_{\sigma}$ is rescaled according to the bond length. We use the
available structural information ~\cite{Ohoyama2001,Onimaru2008} on the
environment of first neighbour C ions around a Dy site. Following
Adachi et al. ~\cite{Adachi2002}, in the structure of the quadrupolar
ordered phase, we add displacement along the c axis of the positions
of the C ions from its positions in the high temperature P4/mbm space
group.

In the excited Hamiltonian we introduce a splitting of the core states
due to the intra-atomic quadrupole interaction $Q_{\bar{J}}$
~\cite{Mulders2006}, $[3 \bar{M}^2 - \bar{J} (\bar{J}+1)]
Q_{\bar{J}}$ , where $Q_{\bar{J}}$ is the product of the 3d
quadrupole moment and the f-electron electric field gradient
experienced by the 3d electrons with $Q_{3/2} / Q_{5/2} =
\frac{7}{3}$. The introduction of this splitting is necessary due to
the difficulty in modeling the crystal field and the interactions that
exists in this system, such as the fact that for the 4f shell the
point charge model is not good, screening, or the effect of the
polarization of the 5d states ~\cite{Staub2000}.

The best fit to the x-ray diffraction and absorption spectra taken in
the AFQ phase ($T = 18$~K) ~\cite{Mulders2006} correspond to the parameters
$Q_{5/2} = 0.2$ eV and $V_{\sigma} = 0.1$ eV. The fitting of the
isotropic absorption and the resonant x-ray diffraction energy profile
at the reflection $(00\frac{1}{2})$ is shown in Fig. 1. In the plot, we
also show the shape that the spectra would have without the core-hole
interaction, i.e. $Q_{\bar{J}}=0$. In the calculated spectra we
consider a Boltzmann average in order to take into account the effect
of the temperature. It is worth noting that, from the fitting to the
experimental spectra we use a positive value of $Q_{5/2} = 0.2$ eV,
while in ~\cite{Mulders2006}, a value of -0.4 eV is used. The change of sign
would mean that the positioning of the harmonic oscillators with
different values of $\bar{M}$ in Fig. 2 of reference~\cite{Mulders2006} would be
reversed. Using a negative value of $Q_{5/2}$ in our calculation, would
lead to a much worse agreement with the diffraction energy profile.

\begin{figure}
\begin{center}
\includegraphics[width=4in,angle=270]{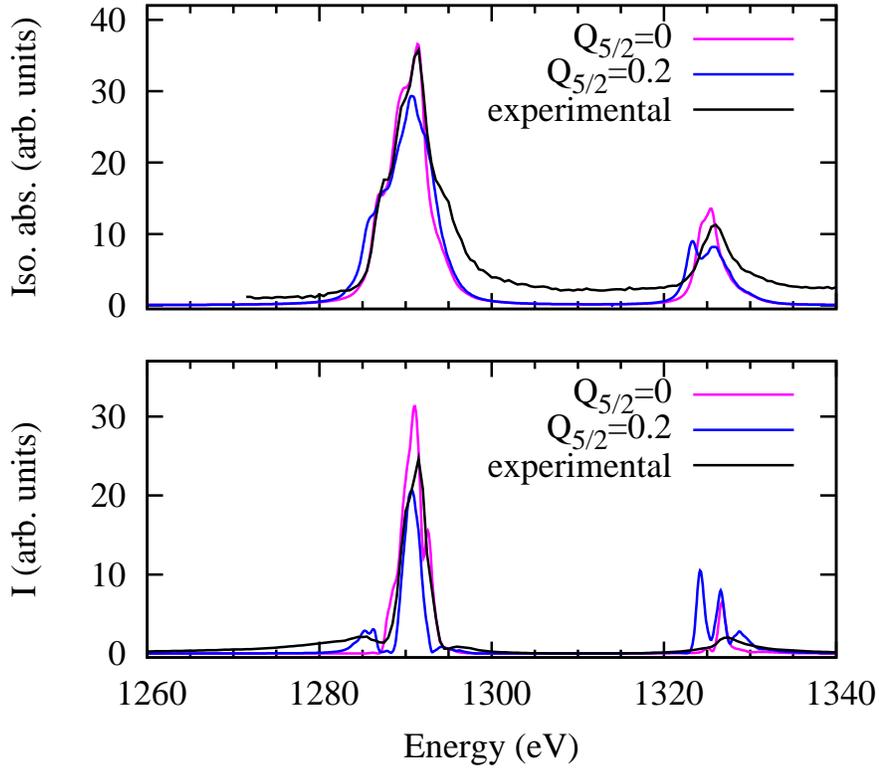}
\end{center}
\caption{\label{Fig1} 
Experimentally measured isotropic absorption and resonant x-ray
diffraction spectra at the $(00\frac{1}{2})$ space group forbidden reflection
and absorption spectra taken in the AFQ phase (T=18 K)
~\cite{Mulders2006} together with the calculated spectra,
corresponding to the parameter $V_{\sigma}$ = 0.1 eV. Calculated
spectra is shown for the cases in which the core-hole interaction is
absent ($Q_{5/2} = 0$) and included ($Q_{5/2} = 0.2$~eV).}
\end{figure}

In order to calculate the spectra below T$_N$ = 15.3 K in the AFM
phase, we calculate the resonant x-ray diffraction spectra introducing
an additional term $\mathbf{S} \cdot \mathbf{H}$ term in the
Hamiltonian, being $\mathbf{S}$ the spin momentum, and $\mathbf{H}$ a
magnetic field wich polarizes the magnetic moment in the ab plane,
forming 23 degrees with respect to the a axis.  The isotropic
absorption does not show any appreciable change when polarizing the
magnetic moment. Fig. 2 shows the diffraction energy profile of the Dy
M$_5$ edge for different values of the spin ($S_{\tilde{z}}$) and
orbital ($L_{\tilde{z}}$) polarization, where {$\tilde{z}$} is the
local anisotropy axis in the ab plane. The Boltzmann average does not
change significantly the spectra, but it reduces notably the values of
$S_{\tilde{z}}$ and $L_{\tilde{z}}$ from the values they would
have at T=0 K. All the curves in Fig. 2 are normalized to have the
same intensity at energies lower than 1282 eV.  Our calculated spectra
shown in Fig. 2 are consistent with the experimentally determined
behavior of the features in the diffraction spectra at E = 1282 eV and
E=1291 eV when lowering the temperature, which is shown in Fig. 1 of
reference~\cite{Mulders2006}.

\begin{figure}
\begin{center}
\includegraphics[width=3in,angle=270]{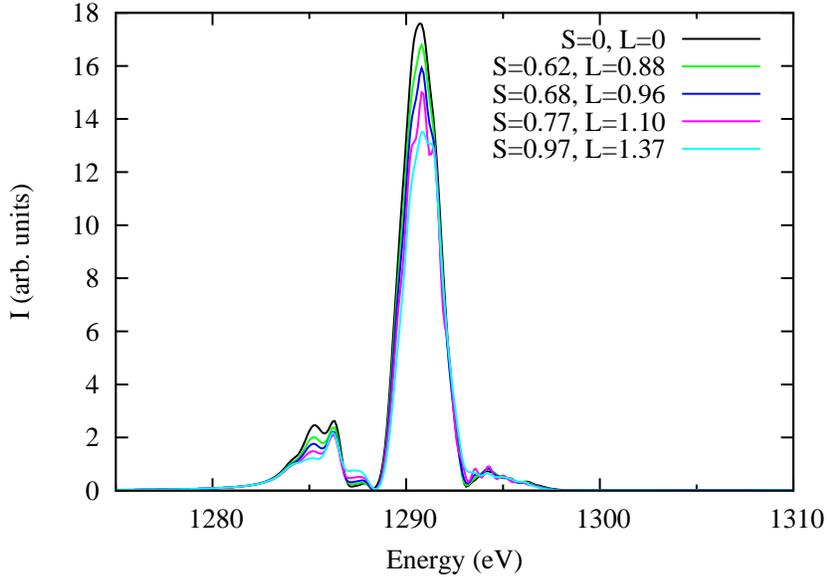}
\end{center}
\caption{\label{Fig2} Evolution of the calculated resonant x-ray diffraction
spectra in the vicinity of the Dy M$_5$ edge at the $(00\frac{1}{2})$ space group
forbidden reflection when the magnetic moment is polarized in the ab
plane forming 23 degrees with respect to the a axis Spectra are shown
for different values of the polarization of the orbital angular
momentum $L_{\tilde{z}}$ and the spin angular momentum $S_{\tilde{z}}$. We normalize all the curves
to have the same intensity at energies lower than 1282 eV.}
\end{figure}

\section{Conclusions}

By using a model taking into account an atomic multiplet Hamiltonian,
crystal field and intra-atomic quadrupolar interaction between the 3d
core-hole and the 4f valence shell we have reproduced the different
features in the experimental isotropic absorption and resonant x-ray
Bragg diffraction energy profile at the Dy M$_4$ and M$_5$ edges in the
antiferroquadrupolar phase of DyB$_2$C$_2$ in terms of a dipolar
transition (4f$^9$ $\longrightarrow$ 3$d^9$ 4$f^{10}$). When, in our model, we vary the
polarization of the magnetic moment of the Dy ion, we observe a
variation of the RXS spectra for different energies around the Dy
M$_5$-edge that agrees with the experimentally observed behaviour of the
measured RXS intensities for different energies when the temperature is
lowered in the antiferromagnetic phase.

\section{Acknowledgements}

We thank J.A. Blanco and V. Scagnoli for useful discussions. One of us,
J.F.R., is grateful to Gobierno del Principado de Asturias for the
financial support from Plan de Ciencia, Tecnologia e Innovacion PCTI de
Asturias 2006-2009.

\end{document}